\documentstyle[twocolumn,prl,aps,epsf]{revtex}
\draft
\begin{document}
\twocolumn[\hsize\textwidth\columnwidth\hsize\csname @twocolumnfalse\endcsname

\title{$d$-electron induced icosahedral growth in strontium clusters }

\author{Tabish Qureshi\cite{email1} and Vijay Kumar\cite{email2}}
\address{Material Science Division,
Indira Gandhi Center for Atomic Research, Kalpakkam-603102, India.}

\maketitle
\begin{abstract}
Strontium clusters of size 2 to 20 atoms have been studied within the
local density approximation using the {\it ab-initio} molecular
dynamics method with a plane wave basis in which the $s$ and $p$
electron wavefunctions are well represented.  The resulting low lying
isomers are optimized using a full-potential linear muffin tin orbital
method which treats the $d$-electrons also on an equal footing. We find
that the $d$ electrons contribute significantly to the binding energy
as the cluster size grows, expedite the onset of metallicity and induce
icosahedral growth. These results are in complete agreement with the
experimental finding of rare-gas like magic clusters of group IIA
elements and suggest that the observation of icosahedral growth in some
transition metal clusters may have a similar origin.

\end{abstract}

\pacs{PACS numbers: 36.40.Cq; 36.40.-c; 71.15.Pd }
]

An interesting question which has intrigued researchers for a long time
is how atoms behave collectively as one goes from molecular state to
the bulk. Small clusters of a variety of materials exhibit quite
different properties as compared to the bulk and are, therefore,
attracting much attention.  It is indeed a challenge of cluster science
to understand the evolution of the atomic structure and the growth of
the physical and chemical properties of matter, from embrionic state to
the bulk.  Clusters of group IIA elements exhibit a non-metal - metal
transition from a weakly bonded dimer to a strongly bonded bulk as the
size grows.  Theoretical studies on $Be$\cite{Be} as well as
$Mg$\cite{Mg} clusters showed their magic numbers to be different from
those of rare gases.  These could be qualitatively understood in terms
of the filling of the electronic shells in a spherical jellium model,
where one assumes that the valence electrons of the atoms move in a
jellium of positively charged ions\cite{jellium}. Magic clusters of
rare gases with 13, 19, 23, 26, 29, 32, 34, ... atoms, however,
correspond to icosahedral growth \cite{raregas} which has been
understood to arise due to a spherically symmetric pairwise interaction
and corresponds to a close packing of hard spheres.  But as one goes
further down the group IIA, one encounters elements $Ca$, $Sr$ and $Ba$
where the unoccupied $d$-states could play an important role in the
growth of the atomic and electronic properties of their clusters.

Currently there is little understanding of the growth and other
properties of metal clusters involving $d$ electrons, in spite of their
importance in several technological applications. Experiments on
clusters of $Ni$\cite{Ni} have provided indirect evidence of
icosahedral growth akin to rare gas clusters, even though bonding in
the latter case is very different.  Also the abundance spectra of
$Eu$\cite{Eu}, $Yb$\cite{Yb}, $Ba$\cite{Ba} and $Sr$\cite{Sr} clusters
show rare gas type magic behavior.  However, the origin of this
seemingly icosahedral growth is not understood. All of these elements
have a $ns^2$ closed shell electronic configuration, and a low-lying
empty valence $(n-1)d$ shell.  Therefore, their clusters can be
expected to exhibit similar behavior. More interestingly, clusters of
$Eu$ have been found to show different abundances depending upon the
nucleation conditions \cite{Eu}.  Low temperature studies on $Eu$
clusters show rare gas type magic clusters whereas experiments with
higher temperature of the carrier gas led to abundances expected from
metal clusters with a free electron like behavior\cite{Eu}. Also,
clusters of coinage metals are found to exhibit features which appear
to be similar to those obtained for alkali metal clusters, but
deviations have also been noted which are believed to suggest an
icosahedral growth and could be due to the presence of $d$
electrons\cite{Cu,Ag}.

In this letter, we show that the $d$-electrons induce icosahedral
growth in metal clusters and explain in particular the magic behavior
observed in the abundance spectra of $Sr$ and $Ba$ clusters.  The
abundance spectrum of small $Sr$ clusters \cite{Sr} is similar to the
one of rare gases but it shows $Sr_{11}$ to be magic which can not be
understood from either the jellium model or the rare gas like picture.
Further, $Sr_{13}$ is surprisingly not magic.  The abundance spectrum
of $Ba$ clusters is similar but $Ba_{13}$ is also magic. To understand
these, we have carried out detailed investigation on $Sr$ clusters.

We have used the Car-Parrinello (CP) \cite{CP} method and simulated
annealing technique to obtain the low lying isomers within the local
density approximation. As a first approximation to the electronic
structure of $Sr$ clusters, we ignored the $d$ states in order to save
the computer time because we expected their occupancy to be small in
clusters.  The ionic potential was  represented by the norm-conserving
pseudopotential of Bachelet {\it et al} \cite{bachelet} in separable
form and the wave functions were expanded in a plane wave basis with a
cut-off of 8 Ry. The simulations were carried out using the
$s$-nonlocality, which treats only the $s$- and $p$- ionic
pseudopotentials accurately. The cluster was placed in an fcc cell of
64 a.u. with periodic boundary conditions. This size of the cell was
large enough so that interaction between the cluster and its periodic
images was negligible. The Brillouin zone was sampled by the $\Gamma$
point and all calculations were performed for the singlet state.  For
clusters with N $\le$ 7, we, however, relaxed a few selected geometries
with steepest descent method. Structures of the low lying isomers were
subsequently optimized using the full potential linear muffin-tin
orbital (FPLMTO) method \cite{lmto} which treats the $s$-, $p$- and
$d$-states on an equal footing. This procedure thus allowed us to study
the role of $d$-orbitals in the bonding of $Sr$ clusters.

\vspace{0.4cm}
\epsfxsize 3in
\epsfbox{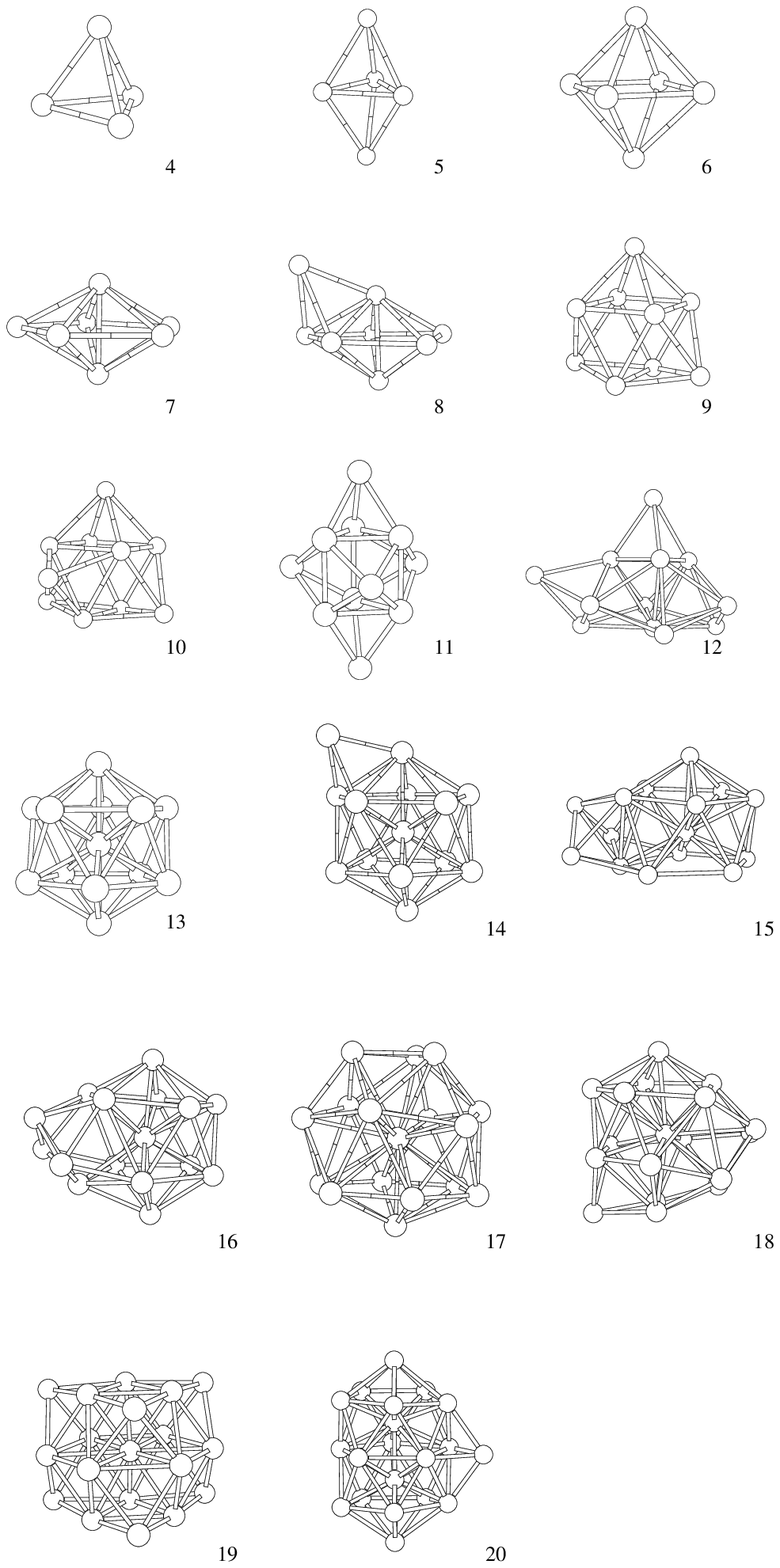}
{\flushleft \sl Figure 1. Minimum energy structures of 
$Sr$ clusters calculated using the CP method. }
\vspace{0.4cm}

The lowest energy structures obtained from the CP method are shown in
Fig. 1. For N $\le$ 11, these are similar to the results obtained
earlier for magnesium clusters\cite{Mg} except for $Sr_6$, which is an
octahedron. $Sr_8$ is a capped pentagonal bipyramid and it is lower in
energy than a bicapped prism as well as a hexagonal bipyramid. For $Sr_9$,
a tricapped prism is lower in energy than a double-capped pentagonal
bipyramid. $Sr_{10}$ is a tetracapped prism whereas in the case of
$Sr_{11}$ all the five faces of a prism get capped and this is lower in
energy than an icosahedron with two atoms missing.

For $Sr_{12}$ simulated annealing yields a tetrahedrally packed
structure to be of lower energy than an icosahedron with one vertex
atom missing. However, for $Sr_{13}$ a regular icosahedron is of lowest
energy, though neither $Be_{13}$, nor $Mg_{13}$ has an icosahedral
structure.  This is due to the fact that the 1$f$ state (of the jellium
model) splits into a 3-fold and a 4-fold state.  The  3-fold state is
completely occupied and it is separated from the unoccupied 4-fold
state by a gap of about 0.5 eV.  The lowest energy structure for
$Sr_{14}$ is a capped icosahedron. $Sr_{15}$  is an incomplete
icosahedron with 3 caps while $Sr_{16}$ is a distorted icosahedron with
3 capping atoms. $Sr_{17}$ is a highly symmetric Frank-Kasper
polyhedron with 16 atoms forming a cage and one atom at the center.
For $Sr_{18}$, the structure has similarity with $Sr_{17}$ and for
$Sr_{19}$ an {\it hcp}-like structure (see Fig. 1) has lower energy
than a double-icosahedron.  $Sr_{20}$ is a double icosahedron with a
cap in the middle.  These results clearly show  that there is no
systematic icosahedral growth if the system is $s-p$ bonded.

\vspace{0.4cm}
\epsfxsize 3in 
\epsfbox{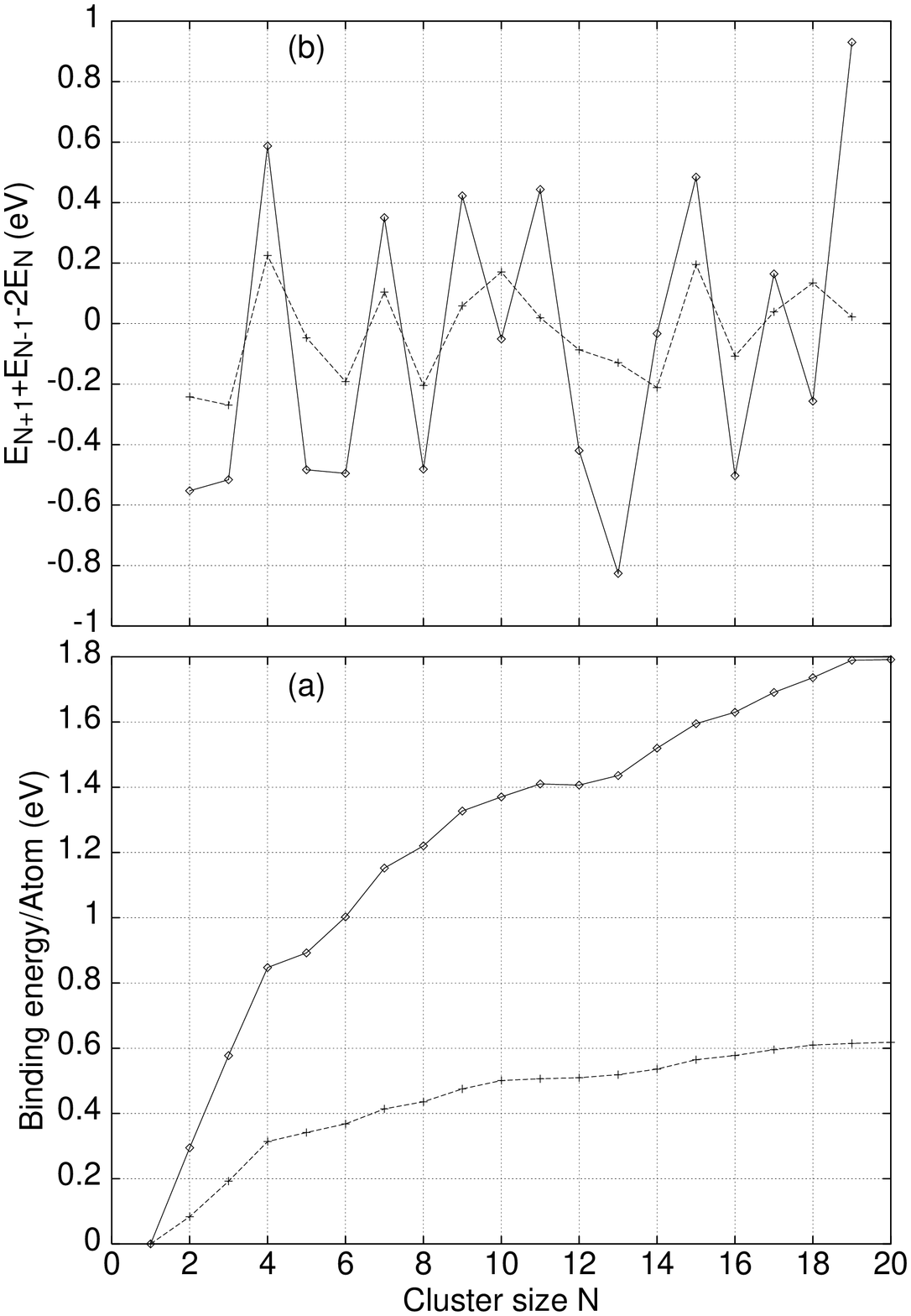} 
{\flushleft \sl Figure 2. (a) Binding energy per atom and (b) second
derivative
of energy for $Sr$ clusters as a function of cluster size. Solid (broken) 
lines denote the results from FPLMTO (CP) calculation.}
\vspace{0.4cm}

The binding energy per atom is shown in Fig. 2. It is noted that the
approach to the bulk value ($\approx$ 2.0 eV/atom \cite{pari}) is
fairly weak.  The second derivative of energy, $\Delta E_N =  E_{N+1} +
E_{N-1} - 2E_N$, $E_N$ being the energy of a $N$ atom cluster, is
significantly positive for $N$ = 4, 7, 10, 15 and 18 favoring a
magic behavior for these clusters (see Fig. 2). An interesting result
to note here is that although the lowest energy structure for $Sr_{13}$
is an icosahedron, it is not magic.  On the other hand, clusters with
10  and 18  atoms  are not magic experimentally. Thus these results do
not agree with experiments on $Sr$ clusters.

In order to investigate the role of $d$-electrons, the low-lying
isomers of all the clusters were optimized using the FPLMTO method,
with $s$, $p$ and $d$ basis orbitals. The resulting lowest energy
structures are shown in Fig. 3 for cases in which the structures are
different from those obtained from the CP method.  It is found that the
$d$ electrons lead to a reduction in the bond lengths in all the
structures investigated.  Most of the larger clusters which did not
have an icosahedral structure to be of lowest energy in the CP method,
now turned out to have very well-defined icosahedral geometries.  Other
structures retained the geometry but there is relaxation in the bond
lengths which have a spread in their values. $Sr_9$ now turned out to
be a pentagonal bipyramid with two caps. Similarly, $Sr_{10}$,
$Sr_{11}$ and $Sr_{12}$ now yielded pentagonal bipyramid with three,
four and five caps respectively. These can also be visualized as
icosahedra with three, two and one atom missing.  From $Sr_{13}$
onwards all clusters are capped icosahedra until one obtains a
double-icosahedron for $Sr_{19}$ with the exception of $Sr_{15}$ which
is a bicapped hexagonal antiprism (see Fig. 3). However, the bicapped
icosahedron, at the binding energy of 1.591 eV/atom, is nearly degenerate
with the minimum energy structure at 1.595 eV/atom. It is clear that now
the icosahedral growth becomes more favorable. Similar structures have
been observed in the experimental studies on $Ni$ clusters\cite{Ni},
and also in theoretical studies using empirical many-body potential for
$Ni$ atoms\cite{Nit}.

The binding energy is substantially larger when the $d$ electrons are also
included (Fig. 2). For $Sr_2$ it is 0.29 eV/atom
in fair agreement with the value of 0.27 eV/atom obtained by Ortiz
and Ballone \cite{Ob}. However, for $Sr_{19}$ the binding energy is
1.75 eV/atom. Therefore, we find a transition from a weak bonding to a
strong metallic bonding as a function of the cluster size.  It is also
interesting to note that the contribution due to the inclusion of the
$d$-states to the binding energy per atom is very small (about 0.18 eV)
for a dimer, but it grows to approximately 1.1 eV for the 19-atom cluster
\cite{note}.  Thus, the $d$-states become increasingly important as the
$sp-d$ hybridization increases with size.  The presence of $d$-states
expedites the approach to the bulk properties as also evidenced from
Fig. 4 by a small highest occupied (HOMO) and the lowest unoccupied
molecular orbital (LUMO) gap ($\approx$ 0.1 eV for $N > 10$).

\vspace{0.4cm}
\epsfxsize 3in
\epsfbox{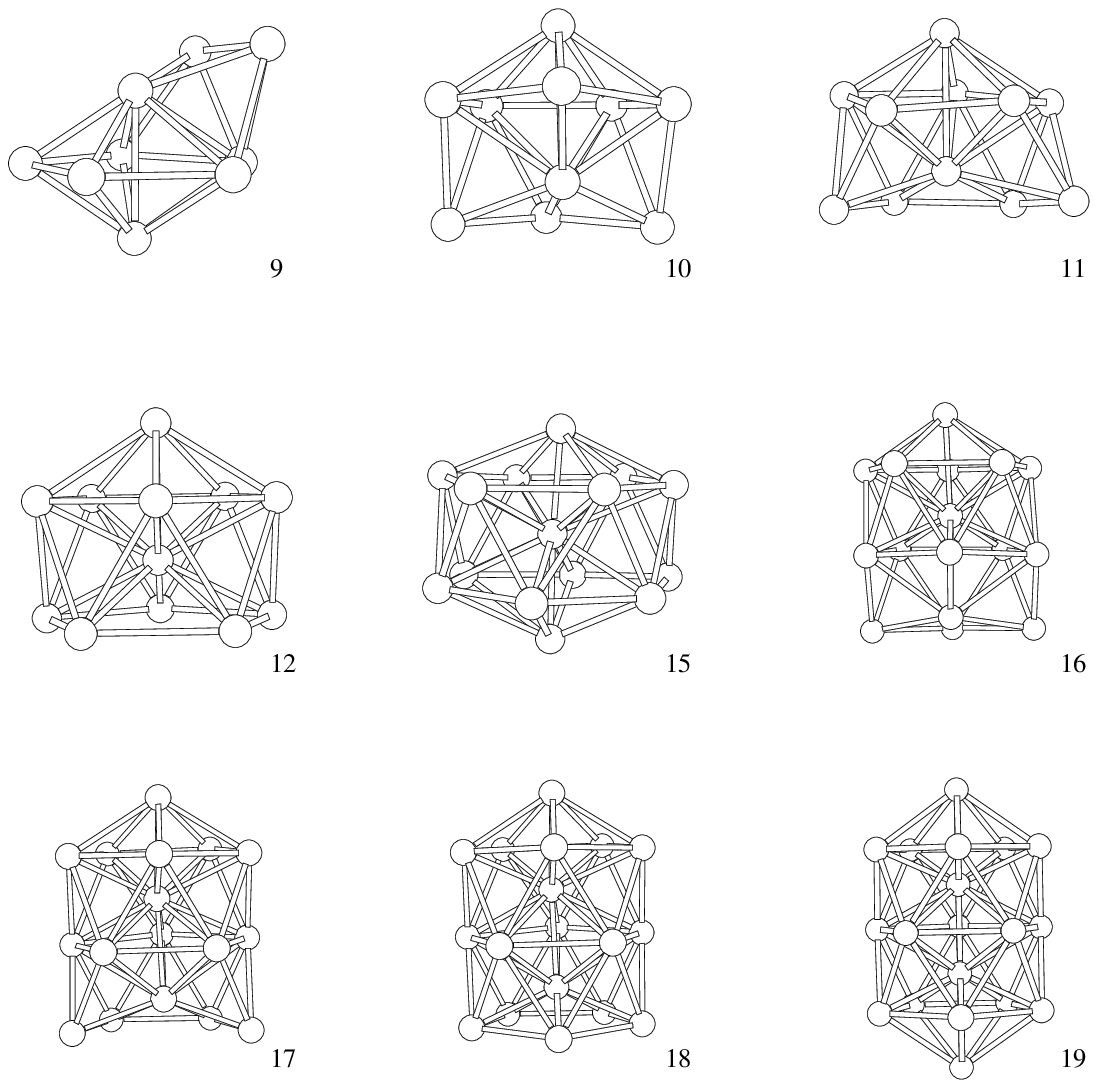}
{\flushleft \sl Figure 3. Lowest energy structures of $Sr$ clusters
obtained 
from FPLMTO method, which differ from those obtained by the CP method. 
The $d$-states favor an icosahedral growth governed by capping the $Sr_7$ 
pentagonal bipyramid.}
\vspace{0.4cm}

We consider our results as compelling evidence that the presence of
unoccupied $d$-states make icosahedral packing more favorable in  $Sr$
clusters. This is the closest packing of atoms in clusters.  Thus the
reason for the occurrence of rare-gas like magic numbers in the
abundance spectrum of $Sr$ clusters is the presence of unoccupied $d$
states close to the $s$ level which participate very significantly in
the bonding.

The second derivative of energy now shows that $Sr_{10}$ is no longer a
magic cluster. On the other hand, $Sr_{11}$  as well as $Sr_{19}$ now
become magic in agreement with the experiment. This substantiates the
generally observed feature that the $s$-$p$ bonded metals fall into the
jellium picture. It shows in addition that the presence of $d$-states
renders the jellium picture inappropriate. Our results with
$d$-electrons also show $Sr_{13}$ not to be magic which is also in
agreement with the experimental data.  We find that for $Sr_{11}$ there
is a significant contraction in the bond length which leads to a
stronger bonding contribution from $d$ states.  Whereas, for $Sr_{13}$,
the bond lengths are slightly longer as compared to the one in
$Sr_{11}$.

Thus our results agree with the experimental finding
for Sr and suggest that the observation of rare gas like magic behavior
for $Ba$, $Yb$ and $Eu$ are also due to the significant contribution of
$d$-states to the   bonding in these clusters. Since $5d$ states get
more populated in $Ba$, $Yb$ and $Eu$, we would expect these to make a
larger contribution which may give rise to a magic behavior for 13-atom
cluster also as is found in the case for $Ni$ clusters.
Our results also suggest that the observation of temperature dependent
abundance spectra could also be due to a change in the $sp-d$
hybridization resulting from thermal expansion in clusters. Note that
for weak contribution from $d$-electrons we find 10- and 18-atom
clusters  to be magic which agrees with the experiments on $Eu$.
Indeed, in our CP simulations on $Sr$ clusters we observed large
increase in the interatomic distance at finite temperatures which
should make such clusters susceptible to structural changes.  We feel
that the icosahedral type of growth observed in coinage metal
clusters \cite{Cu,Ag} beyond a size of 40 atoms may be due to a growing
$sp$-$d$ hybridization.  It is also appropriate to recall here the
structural changes in gold clusters due to electron
irradiation\cite{marks}, which may be a heating effect of radiation. It
is known that the $s-d$ hybridization plays a crucial role in the
stability of the {\em fcc} phase of gold.  Therefore, there is a
critical balance arising from $d$ electrons which transform the
clusters to a closed packed structure.

\vspace{0.4cm}
\epsfxsize 3in 
\epsfbox{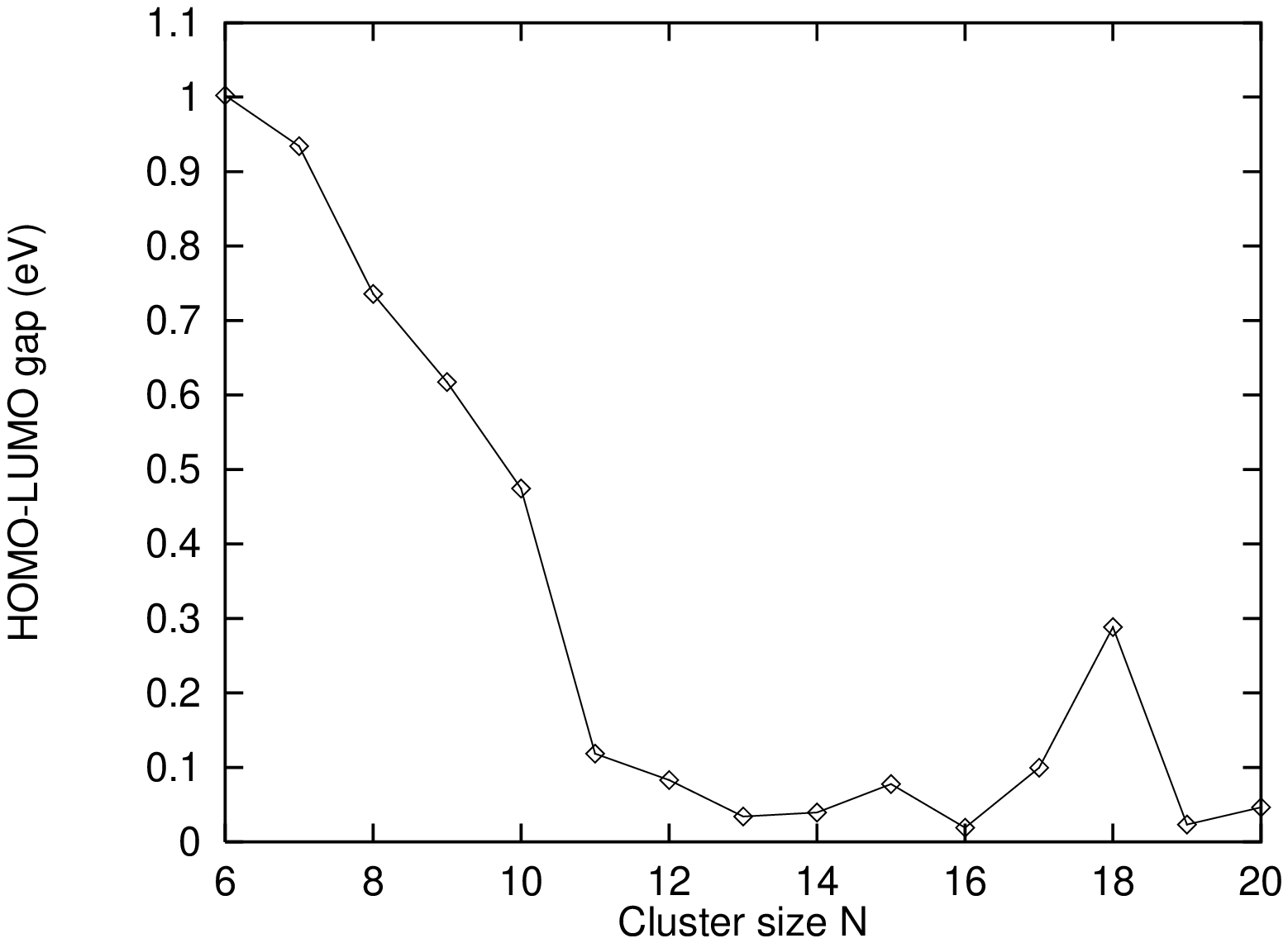}
{\flushleft \sl Figure 4.
The HOMO-LUMO gap for $Sr$ clusters. One can see that it is not correlated
with the peaks of $\Delta E_N$.} 
\vspace{0.4cm}\\

To summarize, we have calculated the structure and electronic
properties of $Sr$ clusters in the range of 2-20 atoms, using the CP
and the FPLMTO methods. We find that the $d$-electrons contribute
substantially to the binding energy and induce icosahedral growth in
these clusters.  The contribution increases from a very small value for
the dimer to a large one for a 20-atom cluster. The calculated magic
numbers are in agreement with the experimental results. These are the
first results where the effect of $d$ electrons has been studied on the
growth of clusters and we hope that our studies would  stimulate
further work in the understanding of the properties of
transition metals clusters.

{\em This work was initiated when VK was a visiting Professor at the
Laboratoire Aime Cotton, CNRS, Universite Paris-Sud, France and he
would like to thank Professors C. Brechignac and P. Cahuzac for
providing the experimental results on Strontium clusters, stimulating
discussions and kind hospitality.  He would also like to thank
Professor C. Colliex and other members of the laboratory for providing
all the help needed to carry out the work there.  We are also
thankful to Dr. G. Pari for his assistance in getting the FPLMTO code
working.}


\begin{references}
\bibitem{note} Note, however, that this includes the changes in energy
of $s$ and $p$ electrons due to hybridization with $d$-states.
Furthermore, the effect of $d$ electrons could be incorporated in the
CP method. However, we expect the difference in energies due to the use
of two different methods to be small as is evident from the work of
Ortiz and Ballone on $Sr_2$ using a plane-wave pseudopotential
method\cite{Ob}.
\end{references}
\end{document}